\documentclass[prb,twocolumn,groupedaddress,superscriptaddress]{revtex4}
\usepackage{graphicx}
\usepackage{amssymb}
\usepackage[usenames]{color}

\begin{document}

\title{The interaction between the magnetic and superconducting order
parameters in a $\mathrm{La_{1.94}Sr_{0.06}CuO_{4}}$ wire }
\author{Meni Shay}
\author{Amit Keren}
\author{Gad Koren}
\author{Amit Kanigel}
\author{Oren Shafir}
\author{Lital Marcipar}
\affiliation{Department of Physics, Technion - Israel Institute of Technology, Haifa
32000, Israel}
\author{Gerard Nieuwenhuys}
\author{Elvezio Morenzoni}
\author{Andreas Suter}
\author{Thomas Prokscha}
\affiliation{Paul Scherrer Institute, CH 5232 Villigen PSI, Switzerland}
\author{Moshe Dubman}
\affiliation{Paul Scherrer Institute, CH 5232 Villigen PSI, Switzerland}
\affiliation{Institut f\"{u}r Physik der Kondensierten Materie,TU Braunschweig,D-38106
Braunschweig,Germany}
\author{Daniel Podolsky}
\affiliation{Department of Physics, University of Toronto, Toronto, Ontario, Canada M5S
1A7}
\date{\today }

\begin{abstract}
We investigate the coupling between the magnetic and superconducting order
parameters in an 8~m long meander line (\textquotedblleft wire") made of a $%
\mathrm{La_{1.94}Sr_{0.06}CuO_{4}}$ film with a cross section of $0.5\times
100$~ $\mathrm{\mu m^{2}}$. The magnetic order parameter is determined using
the Low-Energy muon spin relaxation technique. The superconducting order
parameter is characterized by transport measurements and modified by high
current density. We find that when the superconducting order parameter is
suppressed by the current, the magnetic transition temperature, $T_{m}$,
increases. The extracted sign and magnitude of the Ginzburg-Landau coupling
constant indicate that the two orders are repulsive, and that our system is
located close to the border between first and second order phase transition.
\end{abstract}

\maketitle

When cuprates are doped their low temperature ordered phase changes from an
antiferromagnetic (AFM) to a superconducting (SC) one. The transition takes
place over a range of doping levels where, at low enough temperatures, the
samples are both superconducting and magnetic \cite%
{Niedermayer98,JulienPRL99,Tranquada}. It is natural to expect phase
separation due to the inhomogenous doping. However local probe such as muon
spin relaxation indicates that the magnetic volume fraction is 100\%,
namely, the magnetic field exists everywhere, even in the SC regions \cite%
{Niedermayer98}. Therefore, the nature of the presence of SC and magnetism
is unclear. Are the two orders coupled, and if yes, what are the sign and
strength of the coupling? What is the order of the transition between the
AFM and SC phases as a function of doping? Is it first order with phase
separation or second order with coexistence?

Here we answer this question by looking at the effect of current $I$ on the
magnetic phase transition temperature, $T_{m}$. A current, on the scale of
the second critical current $I_{c2}$, diminishes the superconducting order
parameter. If the two orders interact, the magnetic order parameter is
expected to react to the current and either increase or decrease depending
on the type of coupling between the two orders. This, in turn, will increase
or decrease $T_{m}$, respectively. Therefore, we map the magnetic phase
transition with and without current. We find that, with current, the
magnetic phase transition temperature increases. This results implies that
the orders are coupled, and that they are repulsive. Analysis based on the
Ginzburg-Landau (GL) model shows that the phase transition is close to the
border between first and second order.

\begin{figure}[tbp]
\includegraphics[width=3.2093in]{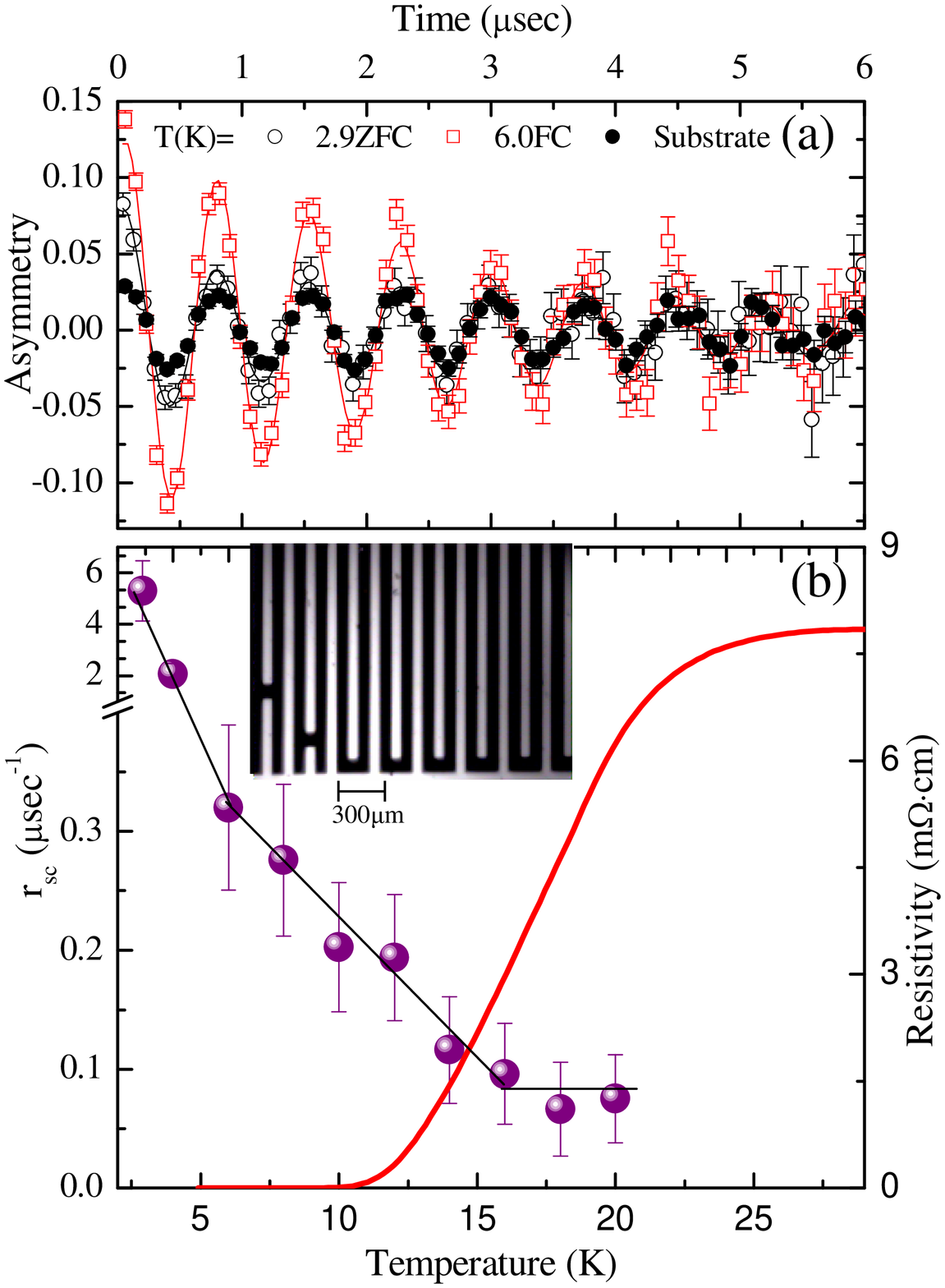} {}
\caption{(color online) Determination of the superconducting volume fraction
and penetration depth (a) $\protect\mu $SR asymmetry under an applied field
of 1 kG in a rotating reference frame at $T=2.9$~K with zero field cooling, $%
T=6.0$~K with field cooling, and at $T=5$~K from the substrate. (b) The
resistivity and muon depolarization rate $r_{sc}$ as a function of
temperature showing $T_{c}$. Below $T_{m}\approx 6$~K the muon relaxation
increases rapidly.}
\label{fig2}
\end{figure}

The experiment is done with an 8~m long wire made of $\mathrm{%
La_{1.94}Sr_{0.06}CuO_{4}}$ film. The film is prepared using laser ablation
deposition on (100) $\mathrm{{LaAlO_{3}}}$ substrate, standard
photolithographic patterning and wet acid etching (0.05\% HCl). The 6\% Sr
doping was chosen since the corresponding bulk material has a $T_{c}\approx
10$~K and $T_{m}\approx 6$~K \cite{Niedermayer98, Panagopoulos2002, Uemura90}%
, which makes both critical temperatures reachable in a standard cryostat.
The cross section of the wire is $0.5$~$\mathrm{{\mu m}\times 100}$~$\mathrm{%
\mu m}$ so that a typical applied current of a few mA is comparable to $%
I_{c2}$. Probing the magnetic properties of such a thin wire is achieved by
using the new low energy muon spin relaxation (LE-$\mu $SR) technique \cite%
{Prokscha2008,em1994prl}. In this technique, the muons are first slowed down
in an Ar moderator where their kinetic energy drops from 4~MeV to 15~eV,
while their initial full polarization is conserved. They are then
electrostatically accelerated to 15~keV and transported in ultra high vacuum
(UHV) to the sample. Four counters collect positrons from the asymmetric
muon decay. One pair of counters is parallel to the initial muon spin
direction and the other pair is perpendicular to it. The muon asymmetry in
these directions is calculated by taking the difference over the sum of the
count for each pair. This asymmetry is proportional to the component of the
muon polarization in each direction. The field the muon experience is either
internal, below $T_{m}$, or external (designated by H), or both. For more
details on $\mu $SR in the presence of superconductivity and magnetism see
Ref.~\onlinecite{SonierRev}. The muons beam spot size has a 15~mm diameter (FWHM).
In order to avoid muons missing the sample, the wire is folded in the form
of a long meandering line covering a disc 3~cm in diameter. The inset of
Fig.~\ref{fig2} shows a magnified image of one corner of the sample.

First, we discuss the sample characterization. In order to verify that the
wire is indeed a bulk superconductor and that the current flows in the bulk
of the wire we performed transverse field LE-$\mu $SR measurements in a
field of $H=1$~kG. Figure~\ref{fig2}(a) depicts the results from the
magnetic phase ($T=2.9$~K) in a rotating reference frame, using zero field
cooling (ZFC). The muons depolarize very quickly and after $3$~$\mathrm{\mu s%
}$ the remaining decay asymmetry is due to muons that have stopped in the
substrate. For comparison, data from a blank substrate, normalized by its
effective area, are also shown. We also present the decay asymmetry in the
pure superconducting phase ($T=6$~K) using field cooling (FC) conditions. In
this case, the muon polarization is lost exponentially versus time at a rate
$r_{sc}$ due to the magnetic field distribution of the vortices in the
superconducting phase. After $6$~$\mathrm{\mu s}$ the polarization reaches
the level of the substrate and the ZFC run, and thus most of the muons are
affected by vortices.

We fit the function:
%\begin{equation}
%Asy(t) =A_{sc}e^{-(r_{n}t)^{2}/2-r_{sc}t}\cos (\omega _{sc}t)+A_{sb}e^{-r_{sb}t}\cos (\omega _{n-sb}t) %+A_{n}e^{-(r_{n}t)^{2}/2}\cos (\omega _{n-sb}t)  \label{eq:Asy}
%\end{equation}%
\begin{eqnarray}
Asy(t) &=&A_{sc}e^{-(r_{n}t)^{2}/2-r_{sc}t}\cos (\omega _{sc}t)+A_{sb}e^{-r_{sb}t}\cos (\omega _{n-sb}t)  \nonumber
\\
&&+A_{n}e^{-(r_{n}t)^{2}/2}\cos
(\omega _{n-sb}t)  \label{eq:Asy}
\end{eqnarray}%
to the muon decay asymmetry at all temperatures. Here $A_{sc}$, $A_{sb}$,
and $A_{n}$ represent the respective contributions from the part of the
meander that turns superconducting upon cooling, the substrate, and the part
of the meander that remains normal upon cooling. $r_{sc}$, $r_{sb}$, and $%
r_{n}$ are the relaxation rates of muons that land in a superconducting,
substrate, and normal material, respectively. $\omega _{n-sb}$ is the
rotation frequencies in the normal material and the substrate (taken to be
equal). $\omega _{sc}$ is the rotation frequency in the superconducting
part. The only parameters that are allowed to vary with $T$ are $r_{sc}$ and
$\omega _{sc}$. The superconducting volume fraction is estimated from $%
A_{sc}/(A_{sc}+A_{n})$ and was found to be $90\pm 5\%$.

Figure~\ref{fig2}(b) shows $r_{sc}$ and the resistivity versus temperature.
The midpoint of the resistivity transition to the superconducting state, and
the onset of $r_{sc}(T)$ occur at $T_{c}=16$~K. The London penetration depth
$\lambda _{ab}$ at $T=7$~K is $500$~nm as estimated from the relation $%
r_{sc}=0.04\gamma _{\mu }\phi _{0}/\lambda _{ab}^{2}$ where $\gamma _{\mu
}/2\pi =13.5$~MHz/kG is the muon gyromagnetic ratio, and $\phi _{0}$ is the
magnetic flux quanta~\cite{Brandt88}. This penetration depth value is
similar to the meander thickness and therefore the current will flow
uniformly in the bulk of the meandering wire.

It is challenging to flow a current in the meander line during a LE-$\mathrm{%
\mu }$SR experiment while keeping its temperature well determined. This
results from the fact that the sample is cooled by a cold finger in a UHV
ambient. Above the first critical current, $I_{c1}$, the superconducting
wire acts as a heater and is not in thermal equilibrium with either the cold
finger or any attached thermometer. Therefore, the wire's temperature can be
measured only by an \emph{a priori} calibration procedure. For this, we
chose to take the V-I curve of the wire at each temperature in a flow
cryostat. In such a cryostat the thermal contact between the wire and a
thermometer, even at high currents, is good. Using this calibration, the
wire acts as its own thermometer. To account for possible drifts in the
calibration we repeated the calibration in the flow cryostat also after the
LE-$\mu $SR experiment. This proved the temperature uncertainty to be
smaller than 0.01 K, namely, when we say that we are comparing two runs with
equal temperatures we mean that we managed to keep the two runs 0.01K away
from each other.

\begin{figure}[]
\includegraphics[width=3.2093in]{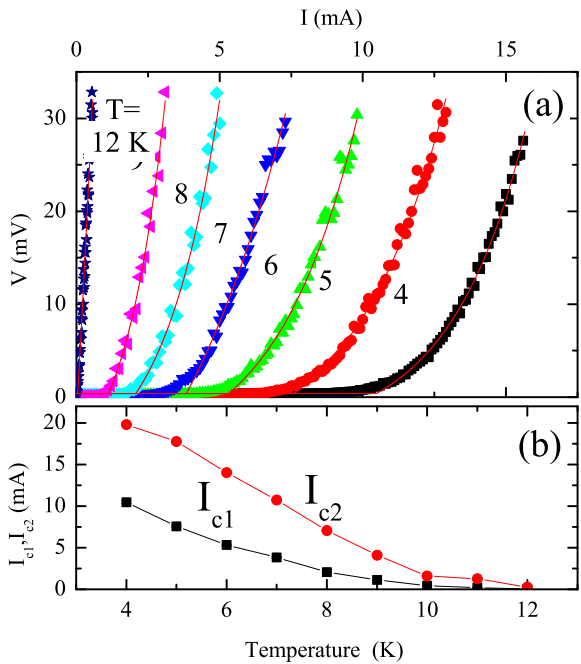} {}
\caption{(color online) Calibration curves used for temperature
determination and for the estimation of $I_{c1}$ and $I_{c2}$.(a) V-I curves
of a short segment of the wire. Similar measurements on the full wire are
used for the temperature calibration. (b) $I_{c1}$ and $I_{c2}$ as a
function of temperature were extracted from the data shown in the top panel.}
\label{fig3}
\end{figure}

Fig.~\ref{fig3}(a) shows several V-I curves recorded at different
temperatures on a short segment (1 cm long) of the wire. These V-I curves
are used for the determination of $I_{c1}$ and $I_{c2}$ which are needed for
the analysis. The curves are fitted to the function $\Theta
(I-I_{c1})e^{k(I-I_{c1})}$, where $\Theta $ is the Heaviside step function.
It is seen in Fig.~\ref{fig3}(a) that, at $T=12$~K, $I_{c1}$ drops to zero
and the 1 cm segment of the wire shows Ohmic behavior with a normal
resistance of $R_{n}=60\Omega $. We estimate $I_{c2}$ using a variation of
the offset criterion \cite{concise}. The exponential dependence of $V$ on $I$
is extrapolated to the value of $I$ that gives a differential resistance
equal to $R_{n}$. The obtained values of both critical currents as a
function of temperature are plotted in Fig.~\ref{fig3}(b).

\begin{figure}[t]
\includegraphics[width=3.2093in]{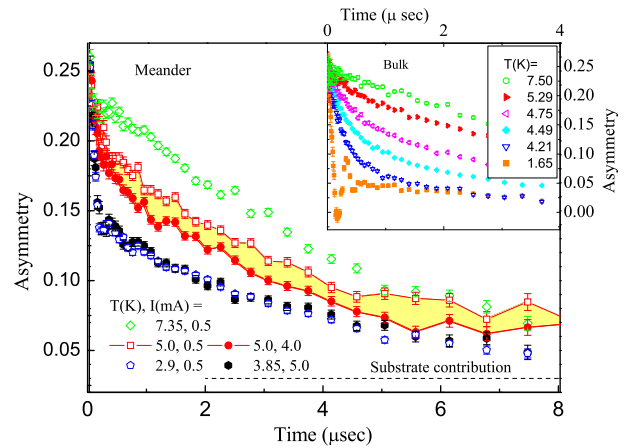}
\caption{(color online) Muon decay asymmetry measurements versus time at low
current (open symbols) and high current (solid symbols). Different colours
represent different temperatures. The area shaded in yellow marks the effect
of the current on the muon decay asymmetry at 5 K. The horizontal line shows
the expected base line from the substrate. The inset shows standard $\protect%
\mu$SR measurements on the bulk powder used for making the film.}
\label{fig4}
\end{figure}

Next, we study the effect of the current on the magnetic order. Figure~\ref%
{fig4} shows raw muon decay asymmetry data from the meander wire at several
temperatures with no external field and in the laboratory frame. The open
symbols represent measurements at low currents (used only for temperature
determination) and the solid symbols are measurements at high currents. At $%
T>T_{m}$, the asymmetry resembles a Gaussian with relatively slow
relaxation, typical of magnetic fields generated by copper nuclear magnetic
moments. As the temperature decreases, there is a clear increase in the muon
spin depolarization rate indicating that the magnetic order has set in. For
comparison, we show in the inset of Fig.~\ref{fig4} standard $\mu $SR
measurements taken with a He flow cryostat on the bulk powder used for
making the film. In this case the measurements could be extended to $T=1.65$
K. We find that the magnetic transition in the wire is very similar to that
of ours and others bulk~samples \cite{Niedermayer98,Uemura90}, having
similar $T_{m}$. In addition, the data in the bulk at low enough
temperatures is typical of the case where muons in the full sample volume
experience frozen magnetism, with spontaneous precession below about $2$~K
with a frequency $f\simeq 3$~MHz, again in agreement with others.

The effect of the current is demonstrated by the $T=5$~K measurement (red
symbols in Fig.~\ref{fig4}). The depolarization of the muons spin is faster
when a higher current is applied. The difference between the two
measurements is emphasized by the shaded area. The change in the asymmetry
line shape caused by the application of current is equivalent to cooling by
about $0.3$~K, although, as mentioned before, the sample temperature is
stable to within $0.01$~K. This effect was observed at several temperatures
along the magnetic transition.

\begin{figure}[t]
\includegraphics[width=3.2093in]{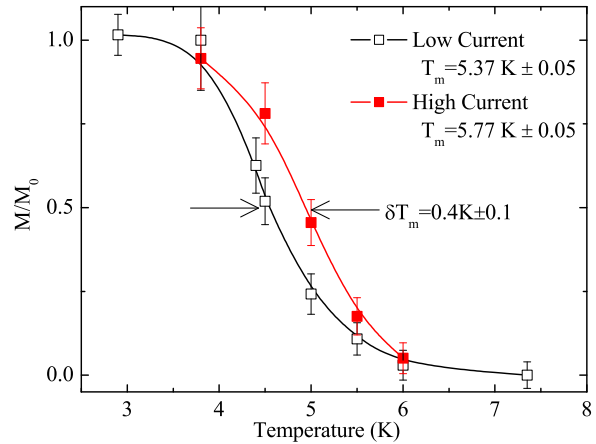}
\caption{(color online) The magnetic phase transition, with and without
current. Solid lines are guide to the eye.}
\label{fig5}
\end{figure}

Above $T_{m}$ and below $4$~K the application of current has no effect on
the asymmetry. This finding is particularly important since, \emph{a priori}%
, the current might affect the muon asymmetry directly by means of the
magnetic field it produces, or by colliding with the muons. However, we
found that once the electronic spins are fully frozen the current does not
change the muon asymmetry indicating that there is no direct current muon
coupling. This is in agreement with calculations showing that the magnetic
field the current produces is very small compared to the internal field.
Similarly, the lack of current effect above $T_{m}$ rules out collisions
between muon and electron charge.

In order to determine the magnetic phase transition temperature, without
assuming a specific spatial field distribution or temporal fluctuation
model, we define the order parameter in a model-free way. At each
temperature the asymmetry as a function of time is averaged to produce $%
\langle Asy\rangle =\frac{1}{t_{m}}\int_{0}^{t_{m}}Asy(t)dt$ where the
measurement time $t_{m}=8$ $\mu \sec $. We expect $\langle Asy\rangle $ to
decrease with increasing magnetic moment size $M(T)$, and therefore defined
\begin{equation}
\frac{M(T)}{M(0)}\equiv \frac{\langle Asy\rangle ^{-1}\left( T\right)
-\langle Asy\rangle ^{-1}\left( \infty \right) }{\langle Asy\rangle
^{-1}\left( 0\right) -\langle Asy\rangle ^{-1}\left( \infty \right) }.
\label{eq:mPol}
\end{equation}%
For $\langle Asy\rangle \left( \infty \right) $ we take the averaged $Asy$
at $T=7.35$~K, which is above the transition. The magnetic phase transition
temperature $T_{m}$ is taken as the onset of the sudden change in $M(T)$.
The magnetic transition is sharp enough that other, model-based, analysis
methods gave indistinguishable $M(T)$. The temperature dependence of $M$
with and without current is presented in Fig.~\ref{fig5}. We find that the
application of a current of about $0.2\cdot I_{c2}(T)$ increases the
magnetic phase transition temperature by $0.4\pm 0.1$~K. This effect means
that the two orders interact repulsively. It is complementary to the effect
of a strong magnetic field on doped samples, where the magnetic order is
enhanced while the superconducting order is suppressed~\cite%
{Katano2000,Lake2002}. However, since current, in contrast to magnetic
field, does not couple directly to spins, the effect presented here is more
simply analyzed. For example, it shows that the enhanced magnetism in the
applied field could be a result of supercurrent in the bulk \cite%
{DemlerPRL01}, and not necessarily due to magnetism in the vortex core \cite%
{HuJCP}.

A simple interpretation of the result can be given in the framework of the
GL model. In this model the free energy density near the critical
temperature $T_{m}$ can be written as $F=-a(T)\left(
1-I^{2}/I_{c2}^{2}\right) |\psi |^{2}+U_{s}|\psi |^{4}-b\left(
T_{m}^{0}-T\right) |\phi |^{2}+U_{m}|\phi |^{4}+2U_{sm}|\phi |^{2}|\psi
|^{2} $ (plus gradient terms) where $\psi $ and $\phi =M/\sqrt{v}\mu _{B}$
are the superconducting and magnetic order parameters respectively, $U_{sm}$ is their coupling constant, $v$ is the
unit cell volume,  $b$ is a dimensionless parameter, $T_{m}^{0}$ is the magnetic phase transition
temperature for $|\psi |^{2}=0$, $a(T)$, $U_{s}$ and $U_{m}$ are the
standard GL parameters. All the parameters can be experimentally determined~%
\cite{Huang,DeGennes}: ${a(T)}={\hbar ^{2}/2m}^{{\ast }}\xi ^{2}$ where $\xi =2~$nm
is the superconducting coherence length~\cite{EPL64}; ${\psi _{0}^{2}}%
=m^{\ast }/{4\mu _{0}e^{2}}\lambda ^{2}$ where $\lambda =500$~nm is the
London penetration depth; $U_{s}=a/2\psi _{0}^{2}$ according
to the minimum condition; ${bT}_{m}={\hbar ^{2}/2m\kappa }^{2}$ where ${%
\kappa }=4$~nm is the magnetic coherence length~\cite%
{MagneticCoherencelength1,MagneticCoherencelength2}; the electron mass can
be approximated by the stiffness of the xy model where $\hbar ^{2}/mA=J$, $A$
is the cell area and $J\simeq 10^{3}$~K is the superexchange; from the ratio
of muon oscillation frequency between our sample and pure $\mathrm{{%
La_{2}CuO_{4}}}$~\cite{Magneticmoment} we find a local magnetic moment $%
M=0.33\mu _{B}$ giving $\phi ^{2}=0.33^{2}/v$; $U_{m}=bT_{m}/2\phi _{0}^{2}$
again by the minimum condition.

$U_{sm}$ is obtained from our current dependent measurement (neglecting
gradient terms at this stage). Since $T_{c}$ is higher than $T_{m}$ we do
not expect $|\phi |^{2}$ to affect $|\psi |^{2}$. Therefore $|\psi
(I,T)|^{2}=|\psi (0,T)|^{2}(1-I^{2}/I_{c2}^{2})$. The minimization of $F$
with respect to $|\phi |^{2}$ yields, $|\phi |^{2}={b(T_{m}^{0}-2U_{sm}|\psi
(I,T)|^{2}/b-T)}/{2U_{m}}$. Thus, the measured magnetic transition
temperature is given by $T_{m}=T_{m}^{0}-2U_{sm}|\psi (I,T_{m})|^{2}/b$. We
assume that near $T_{m}$, $\psi ^{2}(0,T)=\psi _{0}^{2}$ where $\psi
_{0}^{2} $ is the ground state value of $\psi ^{2}$. Therefore, the change
in the transition temperature, $\delta T_{m}\equiv T_{m}(I)-T_{m}(0)$,
caused by the current is $\delta T_{m}(I)={2U_{sm}\psi
_{0}^{2}I^{2}/bI_{c2}^{2}}.$ The interesting parameter is
\[
R\equiv \frac{U_{sm}}{\sqrt{U_{s}U_{m}}}=\frac{2e\lambda \xi
MI_{c2}^{2}\delta T_{m}}{\mu _{_{B}}\hbar \kappa I^{2}T_{m}}\sqrt{\frac{J\mu
_{0}}{h}}
\]%
where $h$ is the unit cell height. For $R>1$ the GL model predicts phase
separation and first order phase transition. For $R<1$ the model predicts
coexistence and a second order phase transition. The $R=1$ condition is
essential for SO(5) symmetry~\cite{Demler04}. At $T=5$~K we found that $%
I_{c2}=17$~mA (see Fig.~\ref{fig3}b) and used $I=4$~mA in the LE-$\mu $SR.
This yields a positive $R=1.4$. Although numerical factors can change $R$,
they cannot change its proximity to unity.

In summary, we demonstrated the presence of interaction between the magnetic
and superconducting order parameters and measured its sign and strength. We
find that phase transition at zero temperature from magnetic to
superconducting orders, as a consequence of doping, must be very close to
the boarder between first and second order.

We acknowledge very helpful discussions with Assa Auerbach and Yariv Kafri.
We also thank the PSI team for supporting the $\mu $SR experiments, and for
providing the continuous high quality beam. This work was also funded in
part by the Israeli Science Foundation and the joint German-Israeli DIP
project.

\end{document}